\documentclass[sigconf]{acmart}

\AtBeginDocument{%
  \providecommand\BibTeX{{%
    \normalfont B\kern-0.5em{\scshape i\kern-0.25em b}\kern-0.8em\TeX}}}

\usepackage{amsmath}
\renewcommand{\vec}[1]{\boldsymbol{#1}}

\usepackage{subfigure}
\usepackage{multirow}
\usepackage{bm}
\usepackage{graphicx}
\copyrightyear{2019} 
\acmYear{2019} 
\acmConference[CIKM '19]{The 28th ACM International Conference on Information and Knowledge Management}{November 3--7, 2019}{Beijing, China}
\acmBooktitle{The 28th ACM International Conference on Information and Knowledge Management (CIKM '19), November 3--7, 2019, Beijing, China}
\acmPrice{15.00}
\acmDOI{10.1145/3357384.3358010}
\acmISBN{978-1-4503-6976-3/19/11}

\begin{document}
\fancyhead{}

\title{Rethinking the Item Order in Session-based Recommendation with Graph Neural Networks}

\author{Ruihong Qiu}
\affiliation{%
  \institution{The University of Queensland}
  \city{Brisbane}
  \country{Australia}
}
\email{r.qiu@uq.edu.au}

\author{Jingjing Li}
\affiliation{%
  \institution{University of Electronic Science and Technology of China}
  \city{Chengdu}
  \country{China}
}
\email{lijin117@yeah.net}

\author{Zi Huang}
\affiliation{%
  \institution{The University of Queensland}
  \city{Brisbane}
  \country{Australia}
}
\email{huang@itee.uq.edu.au}

\author{Hongzhi Yin}
\authornote{Corresponding author.}
\affiliation{%
  \institution{The University of Queensland}
  \city{Brisbane}
  \country{Australia}
}
\email{h.yin1@uq.edu.au}

\begin{abstract}
Predicting a user's preference in a short anonymous interaction session instead of long-term history is a challenging problem in the real-life session-based recommendation, e.g., e-commerce and media stream. Recent research of the session-based recommender system mainly focuses on sequential patterns by utilizing the attention mechanism, which is straightforward for the session's natural sequence sorted by time. However, the user's preference is much more complicated than a solely consecutive time pattern in the transition of item choices. In this paper, therefore, we study the item transition pattern by constructing a session graph and propose a novel model which collaboratively considers the sequence order and the latent order in the session graph for a session-based recommender system. We formulate the next item recommendation within the session as a graph classification problem. Specifically, we propose a weighted attention graph layer and a Readout function to learn embeddings of items and sessions for the next item recommendation. Extensive experiments have been conducted on two benchmark E-commerce datasets, \textit{Yoochoose} and \textit{Diginetica}, and the experimental results show that our model outperforms other state-of-the-art methods.
\end{abstract}



\maketitle

\section{Introduction}
Recommender system (RS) is an important tool to precisely advertise interested items to potential users in today's flood of web information. In recent years, the content-based RS~\cite{Pazzani2007ContentBasedRS} and the collaborative filtering RS~\cite{Schafer2007,he2017neural} are two widely used methods because they can effectively approximate the similarity between items while being simple and efficient. However, a clear drawback of these approaches is that a user's recent preference is ignored. In many scenarios, for example, e-commerce, customers may have their current prioritized choices of products over other products because of their recent need. Consequently, this shift of preference will only be shown in recent interactions between users and items~\cite{wang2019survey}. In this case, the content-based RS and the collaborative filtering RS would fail to capture the important change of the user's preference, which leads to useless or even negative recommendations.

In contrast, a session-based RS can deal with the shift of the user's preference by taking a recent session of user-item interactions (e.g., clicks of items within 24 hours) into consideration. However, it is very common that for modern commercial online systems, they do not record the user's long-term history. In order to make use of the user's preference, the session can be regarded as the representation of the recent preference of the current anonymous user~\cite{wang2019survey}. As a result, how to represent a user's preference by extracting representative information from interactions in the session is the essence of the session-based RS.

As a session is a slice of interactions divided by time, it can be naturally represented as a time series sequence. The sequence characteristic is considered as the most important information by recent methods. However, it has some challenging limitations:

\begin{itemize}
    \item The user's preference does not completely depend on the chronology of the sequence. With the prevalence of RNN~\cite{hochreiter1997long} applied on the sequence data, for instance, GRU4REC~\cite{hidasi2015session} and NARM~\cite{li2017neural} mainly model the time order of items and encode these items using a RNN like neural networks. After encoding the item, the representation of the session is a combination of the item features. Such a paradigm of dealing with the session is natural with the original order of items inside the session. However, as mentioned above, the shift of the user's preference within the session indicates that items should not be simply considered as time series. The item transition pattern is more complicated.
    \item The recent approaches~\cite{wu2018session,li2017neural,Liu18STAMP}, which divide the user's preference into the long-term (global) and the short-term (local) preference, are too simple to capture the complicated item transition pattern. These methods choose the last item of the session to stand for the short-term (local) preference and the remaining items for the long-term (global) preference. This setting directly ignores the pattern of item choices, which introduces the bias to the model. NARM~\cite{li2017neural} applies a self-attention on the last item after encoding with a RNN. To alleviate the influence of time order, STAMP~\cite{Liu18STAMP} only utilizes the self-attention mechanism without RNN. SR-GNN~\cite{wu2018session} proposes to use a single layer gated graph neural network~\cite{li2015gated} to learn the representation of items and again, a self-attention on the last item to extract a session level feature. Actually, the self-attention calculates the relative importance of the last single item, which ignores a specific item transition pattern within the session.
\end{itemize}

With the problems mentioned above, it is important to determine the intrinsic order of items within a session. This inherent order is neither the straightforward time order by RNN, nor the complete randomness by the self-attention. In this paper, a model named Full Graph Neural Network (FGNN) is proposed to learn the inherent order of the item transition pattern and compute a session level representation to generate recommendation. To utilize graph neural networks, we build a session graph for every session and formulate the recommendation as a graph classification problem. In order to capture the inherent order of the item transition pattern, which is vital for the item level feature representation, a multiple weighted graph attention layer (WGAT) network is proposed to compute the information flow between items within the session. After obtaining the item representations, the Readout function, which automatically learns to determine an appropriate order, is deployed to aggregate these features. Extensive experiments are conducted on two benchmark e-commerce datasets, the \textit{Yoochoose} dataset from the RecSys Challenge 2015 and the \textit{Diginetica} dataset from CIKM Cup 2016. The experimental results show the superiority of our method in the task of next item recommendation. Our main contributions are summarized as follows:

\begin{itemize}
\item To the best of our knowledge, we are the first to investigate the inherent order of item transition pattern in session-based recommendation. Specifically, we propose a novel FGNN model to perform the next item session-based recommendation based on the inherent order.
\item A novel WGAT model is proposed to serve as the item feature encoder by learning to assign different weights to different neighbors. It helps to effectively convey the information between items.
\item A Readout function is applied to generate appropriate graph level representation for item recommendation. The Readout function can learn the best order of the item transition pattern in the graph.
\item We conduct extensive experiments on two benchmark e-commerce datasets and achieve state-of-the-art results.
\end{itemize}

\section{Related Work}
\label{related-work}
In this section, we first review some related work about the general recommender system (RS) in Section~\ref{rs} and the session-based recommender system (SBRS) in Section~\ref{sbrs}. At last, we will describe graph neural networks (GNN) for the node representation learning and graph classification problems in Section~\ref{gnn}.

\subsection{General Recommender System}
\label{rs}
The most popular method in recent years for the general recommender system is the collaborative filtering (CF), which represents the user interest based on the whole history. For example, the famous shallow method, Matrix Factorization (MF)~\cite{koren2009matrix} factorizes the whole user-item interaction matrix with latent representation for every user and item. With the prevalence of deep learning, neural networks are widely used. Neural collaborative filtering (NCF)~\cite{he2017neural} first proposes to use the multi layer perceptron to approximate the matrix factorization process. More subsequent work extends the incorporation of different deep learning tools, for instance, zero-shot learning and domain adaptation~\cite{Li19from,li2019both}. These methods all depend on the identification of users and the whole record of interactions for every user. However, the user information is anonymous for many modern commercial online systems, which leads to the failure of these CF based algorithms.

\subsection{Session-based Recommender System}
\label{sbrs}
The research on the session-based recommender system (SBRS) is a sub-field of RS. Compared with RS, SBRS takes the user's recent user-item interactions into consideration rather than requiring all historical actions. SBRS is based on the assumption that the recent choice of items can be viewed as the recent preference of a user. 

{\bf Sequential recommendation} is based on the Markov chain model~\cite{shani2005mdp,zimdars2001using}, which learns the dependence of items of a sequence data to predict the next click. Using probabilistic decision-tree models, Zimdars et al.~\cite{zimdars2001using} proposed to encode the state of the item transition pattern. Shani et al.~\cite{shani2005mdp} made use of a Markov
Decision Process (MDP) to compute the probability of recommendation with the transition probability between items.

\begin{figure*}[ht]
    \centering
    \includegraphics[width=\linewidth]{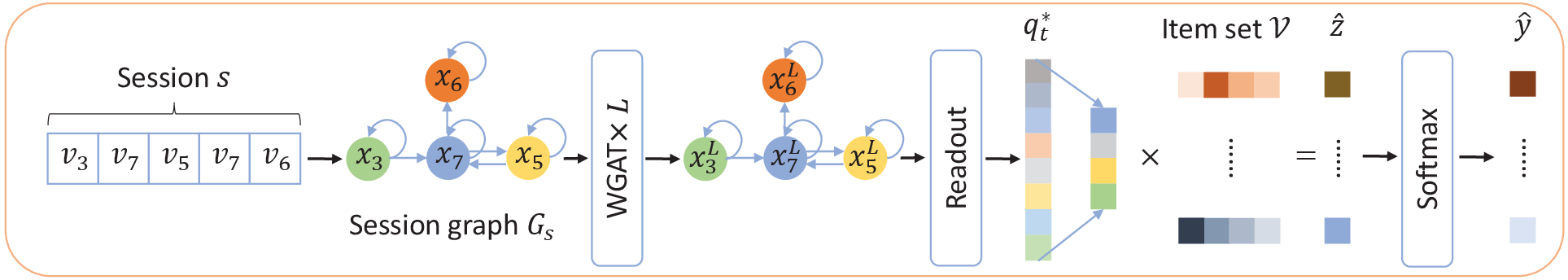}
    \vspace{-0.5cm}
    \caption{Pipeline of FGNN. The input to the model is organized as a session sequence $s$, which is then converted to a session graph $G$ with node features $\vec x$. $L$ layers of WGAT serves as the encoder of node features for $G$. After being processed by WGAT, the session graph now contains different semantic node representations $x^L$ but with the same structure as the input session graph. The Readout function is applied to generate a session embedding based on the learned node features. Compared with other items in the item set $\mathcal{V}$, a recommendation score $\hat{\vec y_i}$ is finally generated.}
    \label{fig:whole}
\end{figure*}

{\bf Deep learning models} are popular recently with the boom of recurrent neural networks~\cite{hochreiter1997long,chung2014empirical,Chen19AIR,Wang16SPORE,Sun2019what}, which is naturally designed for processing sequential data. Hidasi et al.~\cite{hidasi2015session} proposed the GRU4REC, which applies a multi layer GRU~\cite{chung2014empirical} to simply treat the data as time series. Based on the RNN model, some work makes improvements on the architectural choice and the objective function design~\cite{hidasi2018recurrent,tan2016improved}. In addition to RNN, Jannach and Ludewig~\cite{jannach2017recurrent} proposed to use the neighborhood-based method to capture co-occurrence signals. Incorporating content features of items, Tuan and Phuong~\cite{tuan20173d} utilized 3D convolutional neural networks to learn more accurate representations. Wu et al.~\cite{wu2017session} proposed a list-wise deep neural network model to train a ranking model. Some recent work uses the attention mechanism to avoid the time order. NARM~\cite{li2017neural} stacks GRU as the encoder to extract information and then a self-attention layer to assign weight to each hidden state to sum up as the session embedding. To further alleviate the bias introduced by time series, STAMP~\cite{Liu18STAMP} entirely replaces the recurrent encoder with an attention layer. SR-GNN~\cite{wu2018session} applies a gated graph network~\cite{li2015gated} as the item feature encoder and a self-attention layer to aggregate the item features together as the session feature. SSRM~\cite{guo2019streaming} considers a specific user's history sessions and applies the attention mechanism to combine them. Though the attention mechanism can proactively ignore the bias introduced by the time order of interaction, it considers a session as a totally random set.

\subsection{Graph Neural Networks}
\label{gnn}
In recent years, GNN has attracted much interest in the deep learning society. Initially, GNN is applied to the simple situation on directed graphs~\cite{gori2005new,scarselli2009graph}. In recent years, many GNN methods~\cite{kipf2017semi,velickovic2018graph,hamilton2017inductive,li2015gated,xu2018how} work under the mechanism that is similar to message passing network~\cite{gilmer2017neural} to compute the information flow between nodes via edges. Additionally, the graph level feature representation learning is essential for graph level tasks, for example, graph classification and graph isomorphism~\cite{xu2018how,li2019graph}. Set2Set~\cite{vinyals2015order} assigns each node in the graph a descriptor as the order feature and uses this re-defined order to process all nodes. SortPool~\cite{zhang2018end} sorts the nodes based on their learned feature and uses a normal neural network layer to process the sorted nodes. DiffPool~\cite{ying2018hierarchical} designs two sets of GNN for every layer to learn a new dense adjacent matrix for a smaller size of the new densely connected graph.

\section{PRELIMINARIES}
\label{preliminaries}
In this section, we introduce how GNN works on the graph data. Let $G(V,E)$ denote a graph, where $v\in V$ is the node set with node feature vectors $\vec X_v$ and $e\in E$ is the edge set. There are two commonly popular tasks, e.g., \textit{node classification} and \textit{graph classification}. In this work, we focus on graph classification because our purpose is to learn a final embedding for the session rather than single items. For the \textit{graph classification}, given a set of graphs $\{G_1,\ldots,G_N\}\subseteq \mathcal{G}$ and the corresponding labels $\{y_1,\ldots,y_N\}\subseteq \mathcal{Y}$, we aim to learn a representation of the graph $\vec h_G$ to predict the graph label, $y_G=g(\vec h_G)$.

GNN makes use of the structure of the graph and the feature vectors of nodes to learn the representation of nodes or graphs. In recent years, most GNN work by aggregating information from neighboring nodes iteratively. After $k$ iterations of update, the final representations of the nodes capture the structural information as well as the node information within $k$-hop neighbor. The procedure can be formed as
\begin{equation}
    \vec a_v^{(k)}=\text{Agg}(\{\vec h_u^{(k-1)},u\in N(v)\}), \vec h_v^{(k)}=\text{Map}(\vec h_v^{(k-1)},\vec a_v^{(k)}),
\end{equation}
where $\vec h_v^{(k)}$ is the feature vector for node $v$ in the $k$th layer. For the input $\vec h_v^{0}$ to the first layer, the feature vectors $\vec X_v$ are passed in. Agg and Map are two functions that can be defined in a different form. Agg serves as the aggregator to aggregate features of neighboring nodes. A typical characteristic of Agg is permutation invariant. Map is a mapping to transform the self information and the neighboring information to a new feature vector.

For the graph classification, a Readout function aggregates all node features from the final layer of the graph to generate a graph level representation $\vec h_G$:
\begin{equation}
    \vec h_G=\text{Readout}(\{\vec h_v^{(k)},v\in V\}),
\end{equation}
where the Readout function needs to be permutation invariant as well.

\section{Method}
\label{method}
In this section, we describe our FGNN model in detail. Above all, in Section~\ref{problem}, we define the problem and give out the notations used in this paper. The complete pipeline of the calculation is demonstrated in Figure~\ref{fig:whole}. At first, an input session sequence is converted into a session graph (Section~\ref{session-graph}). After obtaining the session graph, an $L$ weighted graph attentional layer (WGAT) model is applied to perform graph convolution among the nodes (Section~\ref{wgat}). Once the node features are learned, a Readout function combines all these features to form a graph level representation $q^*$ (Section~\ref{read-out}). Based on the graph representation, FGNN makes a recommendation by comparing it with the whole item set $\mathcal{V}$ (Section~\ref{rec}). Finally, we describe the way we train our model in Section~\ref{loss}.

\begin{figure}[!ht]
    \centering
    \subfigure[A session graph.]{
    \label{fig:session-graph}
    \includegraphics[width=0.45\linewidth]{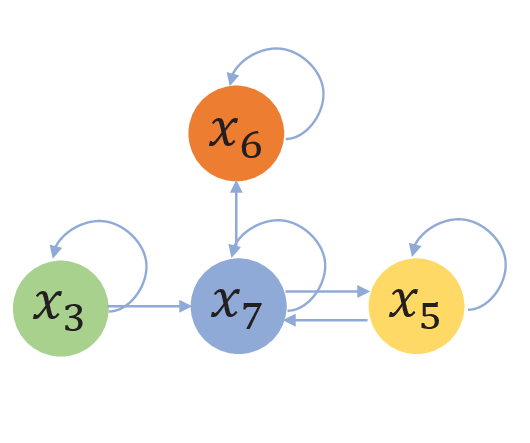}
    }
    \subfigure[The computation for the second-layer feature $\vec x''_6$.]{
    \label{fig:2layer}
    \includegraphics[width=0.49\linewidth]{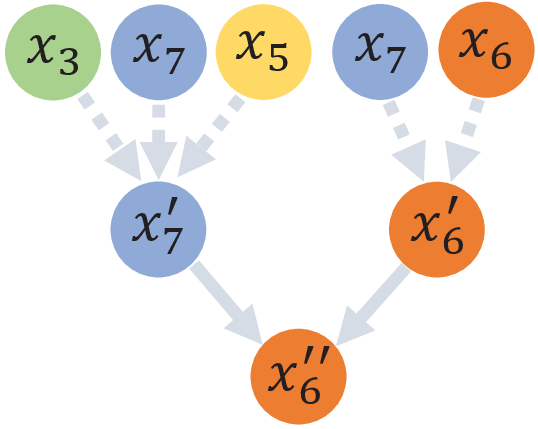}
    }
    \vspace{-0.5cm}
    \caption{An example of how to compute a node representation of a two-layer GNN. The original session sequence is the same as the input in Figure~\ref{fig:whole}.~\subref{fig:session-graph} The session graph is added with self-loop to every node. $\vec x_i$ is the input feature for the corresponding node $v_i$.~\subref{fig:2layer} The computation of the second-layer feature $\vec x''_6$ is based on all the first and second order neighboring nodes of $v_6$. The first order neighbors are $v_7$ and $v_6$ itself. The second order neighbors of $v_6$ are the first order neighbors of $v_7$ and $v_6$, i.e., $v_3$, $v_7$ and $v_6$ for $v_7$, and $v_7$ and $v_6$ for $v_6$.}
\label{fig:propagation}
\end{figure}

\subsection{Problem Definition and Notation}
\label{problem}
The purpose of a session-based recommender system is to predict the next item that matches the user's preference based on the interactions within the session. In the following, we give out the definition of the SBRS problem.

In a SBRS, there is an item set $\mathcal{V}=\{v_1,v_2,v_3,\ldots,v_m\}$, where all items are unique and $m$ denotes the number of items. A session sequence from an anonymous user is defined as a sequential list $\mathcal{S}=[v_{s,1},v_{s,2},v_{s,3},\ldots,v_{s,n}]$, $v_{s,*} \in \mathcal{V}$. $n$ is the length of the session $\mathcal{S}$, which may contain duplicated items, $v_{s,p}=v_{s,q}$, $p, q < n$. The goal of our model is to predict the next item $v_{s,n+1}$ that matches the user's preference the most.

During calculation, for every item $v \in \mathcal{V}$, our model learns a corresponding embedding vector $\vec x \in \mathbb{R}^d$, where $d$ is the dimension of $\vec x$. For each training session, there is a label item $v_{label}$ serving as the target to predict. In order to recommend items based on the given session and the whole item set, our model outputs a probability distribution $\hat{\vec y}$ over $\mathcal{V}$, where the items  with  top-K values in $\hat{\vec y}$ will be the candidates.

\subsection{Session Graph}
\label{session-graph}
As shown in Figure~\ref{fig:whole}, at the first stage, the session sequence is converted into a session graph for the purpose to process each session via GNN. Because of the natural order of the session sequence, we transform it into a weighted directed graph, $G_s=(V_s,E_s)$, $G_s \in \mathcal{G}$, where $\mathcal{G}$ is the set of all session graphs. In the session graph $G_s$, the node set $V_s$ represents all nodes, which are items $v_{s,n}$ from $S$. For every node $\vec v$, the input feature is the initial embedding vector $\vec x$. The edge set $E_s$ represents all directed edges $(v_{s,n-1},v_{s,n},w_{s,(n-1)n})$, where $v_{s,n}$ is the click of the item after $v_{s,n-1}$ in $S$, and $w_{s,(n-1)n}$ is the weight of the edge. The weight of the edge is defined as the frequency of the occurrence of the edge within the session. For the convenience, in the following, we use the nodes in the session graph to stand for the items in the session sequence. For the self-attention used in WGAT introduced in Section~\ref{wgat}, if a node does not contain a self loop, it will be added with a self loop with a weight 1. Based on our observation of our daily life and the datasets, it is common for a user to click two consecutive items for a few times within the session. After converting the session into a graph, the final embedding of $S$ is based on the calculation on this session graph $G_s$.

\subsection{Weighted Graph Attentional Layer}
\label{wgat}
After obtaining the session graph, a GNN is needed to learn embeddings for nodes in a graph, which is the $\text{WGAT}\times L$ part in Figure~\ref{fig:whole}. In recent years, some baseline methods on GNN, for example, GCN~\cite{kipf2017semi} and GAT~\cite{velickovic2018graph}, are demonstrated to be capable of extracting features of the graph. However, most of them are only well-suited for unweighted and undirected graphs. For the session graph, weighted and directed, these baseline methods cannot be directly applied without losing the information carried by the weighted directed edge. Therefore, a suitable graph convolutional layer is needed to effectively convey information between the nodes in the graph.

In this paper, we propose a weighted graph attentional layer (WGAT), which simultaneously incorporates the edge weight when performing the attention aggregation on neighboring nodes. We describe the forward propagation of WGAT in the following. The information propagation procedures are shown in Figure~\ref{fig:propagation}. Figure~\ref{fig:2layer} shows an example of how a two-layer GNN calculates the final representation of the node $v_6$.

The input to a WGAT is a set of node initial features, the item embeddings, $\vec x=\{\vec x_0,\vec x_1,\vec x_2,\ldots \vec x_{n-1}\}$, $\vec x_i \in \mathbb{R}^d$, where $n$ is the number of nodes in the graph, and $d$ is the dimension of the embedding $\vec x_i$. After applying the WGAT, a new set of node features, $\vec x'=\{\vec x'_0,\vec x'_1,\vec x'_2,\ldots \vec x'_{n-1}\}$, $\vec x'_i \in \mathbb{R}^{d'}$, will be given out as the output. Specifically, the input feature vectors $\vec x^0_i$ of the first WGAT layer are generated from an embedding layer, whose input is the one-hot encoding of the item,
\begin{equation}
\label{eq:embed}
    \vec x^0_i=\text{Embed}(v_i),
\end{equation}
where Embed is the embedding layer.

To learn the node representation via the higher order item transition pattern within the graph structure, a self-attention mechanism for every node $i$ is used to aggregate information from its neighboring nodes $\mathcal{N}(i)$, which is defined as the nodes with edges towards the node $i$ (may contain $i$ itself if there is a self-loop edge). Because the size of the session graph is not huge, we can take the entire neighborhood of a node into consideration without any sampling. At the first stage, a self-attention coefficient $e_{ij}$, which determines how importantly the node $j$ will influence the node $i$, is calculated based on $\vec x_i$, $\vec x_j$ and $w_{ij}$,
\begin{equation}
\label{e-ij}
    e_{ij}=\text{Att}(\vec W\vec x_i,\vec W\vec x_j,w_{ij}),
\end{equation}
where Att is a mapping $\text{Att}: \mathbb{R}^{d} \times \mathbb{R}^{d} \times \mathbb{R}^{1} \to \mathbb{R}^{1}$ and $\vec W$ is a shared parameter which performs linear mapping across all nodes. As a matter of fact, the attention of a node $i$ can extend to every node, which is a special case the same as how STAMP makes the attention of the last node of the sequence. Here we restrict the range of the attention within the first order neighborhood of the node $i$ to make use of the inherent structure of the session graph $S$. To compare the importance of different nodes directly, a softmax function is applied to convert the coefficient into a probability form across the neighbors and itself,
\begin{equation}
    \alpha_{ij}=\text{softmax}_j(e_{ij})=\frac{\text{exp}(e_{ij})}{\sum_{k\in \mathcal{N}(i)}\text{exp}(e_{ik})}.
\end{equation}

The choice of $att$ can be diversified. In our experiments, we use an MLP layer with the parameter $\vec W_{att} \in \mathbb{R}^{2d+1}$, followed by a LeakyRelu non-linearity unit with negative input slope $\alpha=0.2$
\begin{equation}
\label{alpha-ij}
    \alpha_{ij}=\frac{\text{exp}(\text{LeakyRelu}(\vec W_{att}[\vec W\vec x_i||\vec W\vec x_j||w_{ij}]))}{\sum_{k\in\mathcal{N}(i)}\text{exp}(\text{LeakyRelu}(\vec W_{att}[\vec W\vec x_i||\vec W\vec x_k||w_{ik}]))},
\end{equation}
where $||$ means concatenation of two vectors.

For every node $i$ in $G_s$, in a WGAT layer, all attention coefficients of their neighbors can be computed as (\ref{alpha-ij}). To utilize these attention coefficients, a linear combination for the corresponding neighbors is applied to update the features of the nodes.
\begin{equation}
\label{1head}
    \vec x'_i=\sigma(\sum\limits_{j\in\mathcal{N}(i)}\alpha_{ij}\vec W\vec x_j),
\end{equation}
where $\sigma$ is a non-linearity unit and in our experiments, we use the ReLU~\cite{nair2010rectified}.

As suggested in previous work~\cite{velickovic2018graph,vaswani2017attention}, the multi-head attention can help to stabilize the training of the self-attention layers. Therefore, we apply the multi-head setting for our WGAT.
\begin{equation}
\label{multihead}
    \vec x'_i=\mathop{\Arrowvert}\limits_{k=1}^K\sigma(\sum\limits_{j\in\mathcal{N}(i)}\alpha^k_{ij}\vec W^k\vec x_j),
\end{equation}
where $K$ is the number of heads and for every head, there is a different set of parameters. $\Arrowvert$ in (\ref{multihead}) stands for the concatenation of all heads. As a result, after the calculation of (\ref{multihead}), $\vec x'_i \in \mathbb{R}^{K{d'}}$.

Specifically, if we stack multiple WGAT layers, the final nodes feature will be shaped as $\mathbb{R}^{K{d'}}$ as well. However, what we expect is $\mathbb{R}^{d'}$. Consequently, we calculate the mean over all heads of the attention results.
\begin{equation}
\label{mean}
    \vec x'_i=\sigma(\frac{1}{K}\sum\limits^K_{k=1}\sum\limits_{j\in\mathcal{N}(i)}\alpha^k_{ij}\vec W^k\vec x_j).
\end{equation}

Once the forward propagation of multiple WGAT layers has finished, we obtain the final feature vector of all nodes, which is the item level embeddings. These embeddings will serve as the input of the session embedding computation stage that we detail below.

\subsection{Readout Function}
\label{read-out}
A Readout function aims to give out a representation of the whole graph based on the node features after the forward computation of the GNN layers. As we introduce above, the Readout function needs to learn the order of the item transition pattern to avoid the bias of the time order and the inaccuracy of the self-attention on the last input item. For the convenience, some algorithms use simple permutation invariant operations for example, $Mean, Max$ or $Sum$ over all node features. Though clearly, the methods mentioned above are simple and perfectly do not violate the constraints of the permutation invariance, they can not provide a sufficient model capacity for learning a representative session embedding for the session graph. In contrast, Set2Set~\cite{vinyals2015order} is a graph level feature extractor which learns a query vector indicating the order of reading from the memory for an undirected graph. We modify this method to suit the setting of the session graphs. The computation procedures are as follows:
\begin{equation}
\label{set2set}
    \vec q_t=\text{GRU}(\vec q^*_{t-1}),
\end{equation}
\begin{equation}
    e_{i,t}=f(\vec x_i,\vec q_t),
\end{equation}
\begin{equation}
    a_{i,t}=\frac{\text{exp}(e_{i,t})}{\sum_j\text{exp}(e_{j,t})},
\end{equation}
\begin{equation}
    \vec r_t=\sum\limits_ia_{i,t}\vec x_i,
\end{equation}
\begin{equation}
\label{q*t}
    \vec q^*_t=\vec q_t\Arrowvert \vec r_t,
\end{equation}
where $i$ indexes node $i$ in the session graph $G_s$, $\vec q_t$, $\vec q_t \in \mathbb{R}^d$, is a query vector which can be seen as the order to read $\vec r_t\in \mathbb{R}^d$ from the memory and GRU is the gated recurrent unit, which at the first step takes no input and at the following steps, takes the former output $\vec q^*_{t-1}\in \mathbb{R}^{2d}$. $f$ calculates the attention coefficient $e_{i,t}$ between the embedding of every node $\vec x_i$ and the query vector $\vec q_t$. $a_{i,t}$ is the probability form of $e_{i,t}$ after applying a softmax function over $e_{i,t}$, which is then used to a linear combination on the node embeddings $\vec x_i$. The final output $\vec q^*_t$ of one forward computation of the Readout function is the concatenation of $\vec q_t$ and $\vec r_t$.

Based on all node embeddings for a session graph, we use Equation~\ref{set2set}$\sim$\ref{q*t} to obtain a graph level embedding which contains a query vector $\vec q_t$ in addition to the semantic embedding vector $\vec r_t$. The query vector $\vec q_t$ controls what to read from the node embeddings, which actually provides an order to process all nodes if we recursively apply the Readout function.

\subsection{Recommendation}
\label{rec}
Once the graph level embedding $\vec q^*_t$ is obtained, we can use it to make a recommendation by computing a score vector $\hat{\vec z}$ for every item over the whole item set $\mathcal{V}$ with the their initial embeddings in the matrix form,
\begin{equation}
\label{z}
    \hat{\vec z}=(\vec W_{out}\vec q^*_t)^T \vec X^0,
\end{equation}
where $\vec W_{out} \in \mathbb{R}^{d\times2d}$ is a parameter that performs a linear mapping on the graph embedding $\vec q^*_t$, the $T$ means the transformation on a matrix, and $\vec X^0$ is from Equation~\ref{eq:embed}.

For every item in the item set $\mathcal{V}$, we can calculate a recommendation score and combine them together, we obtain a score vector $\hat{\vec z}$. Furthermore, we apply a softmax function over $\hat{\vec z}$ to transform it into the probability distribution form $\hat{\vec y}$,
\begin{equation}
\label{y}
    \hat{\vec y}=\text{softmax}(\hat{\vec z}).
\end{equation}

For the top-K recommendation, it is simple to choose the highest K probabilities over all items based on $\hat{\vec y}$.

\subsection{Objective Function}
\label{loss}
Since we already have the recommendation probability of a session, we can use the label item $v_{label}$ to train our model with the supervised learning method.

As mentioned above, we formulate the recommendation as a graph level classification problem. Consequently, we apply the multi-class cross entropy loss between $\hat{\vec y}$ and the one-hot encoding of $v_{label}$ as the objective function. For a batch of training sessions, we can have
\begin{equation}
    L=-\sum\limits_{i=1}^{l}\text{one-hot}(v_{label,i})\text{log}(\hat{\vec y_i}),
\end{equation}
where $l$ is the batch size we use in the optimizer.

In the end, we use the Back-Propagation Through Time (BPTT) algorithm to train the whole FGNN model.

\begin{table*}[!ht]
    \centering
    \caption{Statistic details of datasets.}
    \begin{tabular}{lccccc}
         \toprule
         Dataset&all the clicks&train sessions&test sessions&all the items&avg.length\\
         \midrule
         Yoochoose1/64&557248&369859&55898&16766&6.16\\
         Yoochoose1/4&8326407&5917746&55898&29618&5.71\\
         Diginetica&982961&719470&60858&43097&5.12\\
         \bottomrule
    \end{tabular}
    \label{datasets}
\end{table*}

\section{Experiments}
\label{exp-init}

In this section, we conduct experiments with the purpose to prove the efficacy of our proposed FGNN model by answering the following research questions:
\begin{itemize}
    \item \textbf{RQ1} Does the FGNN outperform other state-of-the-art SBRS methods? (in Section~\ref{rq1})
    \item \textbf{RQ2} How does the WGAT work for the session-based recommendation problem? (in Section~\ref{rq2})
    \item \textbf{RQ3} How does the Readout function work differently from other graph level embedding methods? (in Section~\ref{rq3})
\end{itemize}

In the following, we first describe the details of the basic setting of the experiments and afterwards, we answer the questions above by showing the results.

\subsection{Datasets}
\label{datasets}
We choose two representative benchmark e-commerce datasets, i.e., \textit{Yoochoose} and \textit{Diginetica}, to evaluate our model.
\begin{itemize}
    \item \textit{Yoochoose} is used as a challenge dataset for RecSys Challenge 2015 \footnote{https://2015.recsyschallenge.com/challenge.html}. It is obtained by recording click-streams from an E-commerce website within 6 months.
    \item \textit{Diginetica} is used as a challenge dataset for CIKM cup 2016 \footnote{http://cikm2016.cs.iupui.edu/cikm-cup/}. It contains the transaction data which is suitable for session-based recommendation.
\end{itemize}

For the fairness and the convenience of comparison, we follow~\cite{li2017neural,Liu18STAMP,wu2018session} to filter out sessions of length 1 and items which occur less than 5 times in each dataset respectively. After the preprocessing step, there are 7,981,580 sessions and 37,483 items remaining in \textit{Yoochoose} dataset, while 204,771 sessions and 43097 items in \textit{Diginetica} dataset. Similar to~\cite{wu2018session,tan2016improved}, we split a session of length $n$ into $n-1$ partial sessions of length ranging from $2$ to $n$ to augment the datasets. For the partial session of length $i$ in the session $S$, it is defined as $[v_{s,0},\ldots,v_{s,i-1}]$ with the last item $v_{s,i-1}$ as $v_{label}$. Following~\cite{li2017neural,Liu18STAMP,wu2018session}, for the \textit{Yoochoose} dataset, the most recent portions $1/64$ and $1/4$ of the training sequence are used as two split datasets respectively. Statistical details of all datasets are shown in Table~\ref{datasets}.

\subsection{Baselines}
\label{baselines}
In order to prove the advantage of our proposed FGNN model, we compare FGNN with the following representative methods:
\begin{itemize}
    \item \textbf{POP} always recommends the most popular items in the whole training set, which serves as a strong baseline in some situations although it is simple.
    \item \textbf{S-POP} always recommends the most popular items for the current session.
    \item \textbf{Item-KNN}~\cite{sarwar2001item} computes the similarity of items by the cosine distance of two item vectors in sessions. Regularization is also introduced to avoid the rare high similarities for unvisited items.
    \item\textbf{BPR-MF}~\cite{rendle2009bpr} proposes a BPR objective function which calculates a pairwise ranking loss. Following~\cite{li2017neural}, Matrix Factorization is modified to session-based recommendation by using mean latent vectors of items in a session.
    \item \textbf{FPMC}~\cite{rendle2010factorizing} is a hybrid model for the next-basket recommendation and it achieves state-of-the-art results. For anonymous session-based recommendation, following~\cite{li2017neural}, we omit the user feature directly because of the unavailability.
    \item \textbf{GRU4REC}~\cite{hidasi2015session} stacks multiple GRU layers to encode the session sequence into a final state. It also applies a ranking loss to train the model.
    \item \textbf{NARM}~\cite{li2017neural} extends to use an attention layer to combine encoded states of RNN, which enables the model to explicitly emphasize on the more important parts of the input.
    \item \textbf{STAMP}~\cite{Liu18STAMP} uses attention layers to replace all RNN encoders in previous work to even make the model more powerful by fully relying on the self-attention of the last item in a sequence.
    \item \textbf{SR-GNN}~\cite{wu2018session} applies a gated graph convolutional layer~\cite{li2015gated} to obtain item embeddings, followed by a self-attention of the last item as \textbf{STAMP} does to compute the sequence level embeddings.
\end{itemize}

\subsection{Evaluation Metrics}
\label{eval}
At a time, a recommender system can give out a few recommended items and a user would choose the first few of them. To keep the same setting as previous baselines, we mainly choose to use top-20 items to evaluate a recommender system and specifically, two metrics, i.e., \textbf{R@20} and \textbf{MRR@20}. For more detailed comparison, top-5 and top-10 results are considered as well.

\begin{itemize}
    \item \textbf{R@K} (Recall calculated over top-K items). The R@K score is the primary metric that calculates the proportion of test cases which recommend the correct item in a top K position in a ranking list,
    \begin{equation}
        \text{R@K}=\frac{n_{hit}}{N},
    \end{equation}
    where $N$ represents the number of test sequences $S_{test}$ in the dataset and $n_{hit}$ counts the number that the desired items are in the top K position in the ranking list, which is named the $hit$.
    \item \textbf{MRR@K} (Mean Reciprocal Rank calculated over top-K items). The reciprocal is set to $0$ when the desired items are not in the top K position and the calculation is as follows,
    \begin{equation}
        \text{MRR@K}=\frac{1}{N}\sum\limits_{v_{label}\in S_{test}}\frac{1}{Rank(v_{label})}.
    \end{equation}
    The MRR is a normalized ranking of $hit$, the higher the score, the better the quality of the recommendation because it indicates a higher ranking position of the desired item.
\end{itemize}

\subsection{Experiments Setting}
\label{exp-set}
We apply a three layer WGAT and each with eight heads as our node representation encoder and a three processing steps of our Readout function. The size of the feature vectors of the items are set to 100 for every layer including the initial embedding layer. All parameters of the FGNN are initialized using a Gaussian distribution with a mean of 0 and a standard deviation of 0.1 except for the GRU unit in the Readout function, which is initialized using the orthogonal initialization~\cite{saxe2013exact} because of its performance on RNN-like units. We use the Adam optimizer with the initial learning rate $1e-3$ and the linear schedule decay rate of 0.1 for every 3 epochs. The batch size for mini-batch optimization is 100 and we set an L2 regularization to $1e-5$ to avoid overfitting.

\subsection{Comparison with Baseline Methods (RQ1)}
\label{rq1}
To demonstrate the overall performance of FGNN, we compare it with the baseline methods mentioned in Section~\ref{baselines} by evaluating the R@20 and MRR@20 scores. The overall results are presented in Table~\ref{all-baseline} with respect to all baseline methods and our proposed FGNN model. Due to the insufficient memory of hardware, we can not initialize FPMC on \textit{Yoochoose1/4} as~\cite{li2017neural}, which is not reported in Table~\ref{all-baseline}. For more detailed comparisons, in Table~\ref{tab:5and10}, we present the results of the most recent state-of-the-art methods for the dataset \textit{Yoochoose1/64} when $K=5$ and $10$.

\subsubsection{General Comparison by P20 and MRR20}
\label{sec:gen-com}
FGNN utilizes the multi layers of WGAT to easily convey the semantic and structural information between items within the session graph and applies the Readout function to decide the relative significance as the order of nodes in the graph to make the recommendation. According to the results reported in Table~\ref{all-baseline}, obviously, the proposed FGNN model outperforms all the baseline methods on all three datasets for both metrics, R@20 and MRR@20. It is proved that our method achieves state-of-the-art performance on benchmark datasets. We also substitute the two key components, WGAT and the Readout function, with gated graph networks (FGNN-Gated) and the self-attention (FGNN-ATT-OUT) used by previous methods. Both of the variants perform better than previous models, which demonstrates the efficacy of the proposed WGAT and the Readout function respectively.

\begin{table}[!ht]
    \centering
    \caption{Performance compared with other baselines.}
    \scalebox{0.8}{
    \begin{tabular}{ccccccc}
         \toprule
         \multirow{2}*{Method}&\multicolumn{2}{c}{\textit{Yoochoose1/64}}&\multicolumn{2}{c}{\textit{Yoochoose1/4}}&\multicolumn{2}{c}{\textit{Diginetica}}\\
         &R@20&MRR@20&R@20&MRR@20&R@20&MRR@20\\
         \midrule
         POP&6.71&1.65&1.33&0.30&0.89&0.20\\
         S-POP&30.44&18.35&27.08&17.75&21.06&13.68\\
         Item-KNN&51.60&21.81&52.31&21.70&35.75&11.57\\
         BPR-MF&31.31&12.08&3.40&1.57&5.24&1.98\\
         FPMC&45.62&15.01&-&-&26.53&6.95\\
         GRU4REC&60.64&22.89&59.53&22.60&29.45&8.33\\
         NARM&68.32&28.63&69.73&29.23&49.70&16.17\\
         STAMP&68.74&29.67&70.44&30.00&45.64&14.32\\
         SR-GNN&70.57&30.94&71.36&31.89&50.73&17.59\\
         \midrule
         (ours)\\
         FGNN-Gated&70.85&31.05&71.5&32.17&51.03&17.86\\
         FGNN-ATT-OUT&70.74&31.16&71.68&32.26&50.97&18.02\\
         FGNN&$\bm{71.12}$&$\bm{31.68}$&$\bm{71.97}$&$\bm{32.54}$&$\bm{51.36}$&$\bm{18.47}$\\
         \bottomrule
    \end{tabular}}
    \label{all-baseline}
\end{table}

\begin{table}[!ht]
    \centering
    \caption{Performance when $K=5$ and $10$ for \textit{Yoochoose1/64}.}
    \scalebox{0.8}{
    \begin{tabular}{ccccc}
         \toprule
         \multirow{2}*{Method}&\multicolumn{4}{c}{\textit{Yoochoose1/64}}\\
         &R@5&MRR@5&R@10&MRR@10\\
         \midrule
         NARM&44.34&26.21&57.50&27.97\\
         STAMP&45.69&27.26&58.07&28.92\\
         SR-GNN&47.42&28.41&60.21&30.13\\
         FGNN&$\bm{48.23}$&$\bm{29.16}$&$\bm{60.97}$&$\bm{30.85}$\\
         \bottomrule
    \end{tabular}}
    \label{tab:5and10}
\end{table}

Compared with those traditional algorithms, e.g., POP and S-POP, because of their simple intuition to recommend items based on the frequency of appearance, they perform far worse than FGNN. They tend to recommend fixed items, which leads to the failure of capturing the characteristics of different items and sessions. Taking BPR-MF and FPMC into consideration, which omits the session setting when recommending items, we can see that S-POP can defeat these methods as well because S-POP makes use of the session context information. Item-KNN achieves the best results among the traditional methods, though it only calculates the similarity between items without considering sequential information. At the even worse situation when the dataset is large, methods relying on the whole item set undoubtedly fail to scale well. All methods above achieve relatively poor results compared with the recent neural-network-based methods, which fully model the user's preference in the session sequence.

\begin{figure*}
    \centering
    \subfigure[R@20 index.]{
    \label{fig:p20-wgat}
    \includegraphics[width=0.48\linewidth]{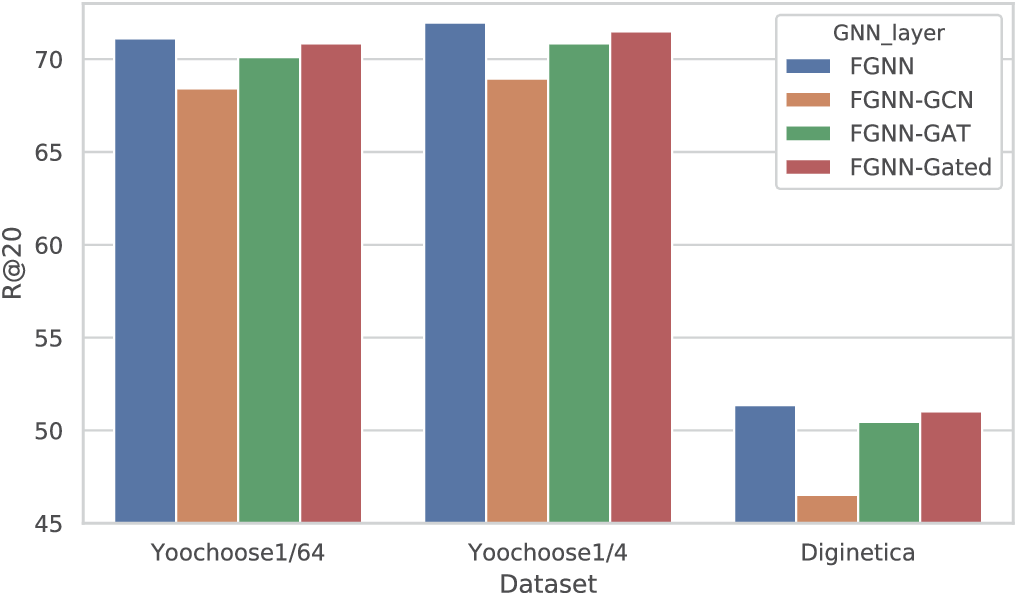}
    }
    \subfigure[MRR@20 index.]{
    \label{fig:mrr20-wgat}
    \includegraphics[width=0.48\linewidth]{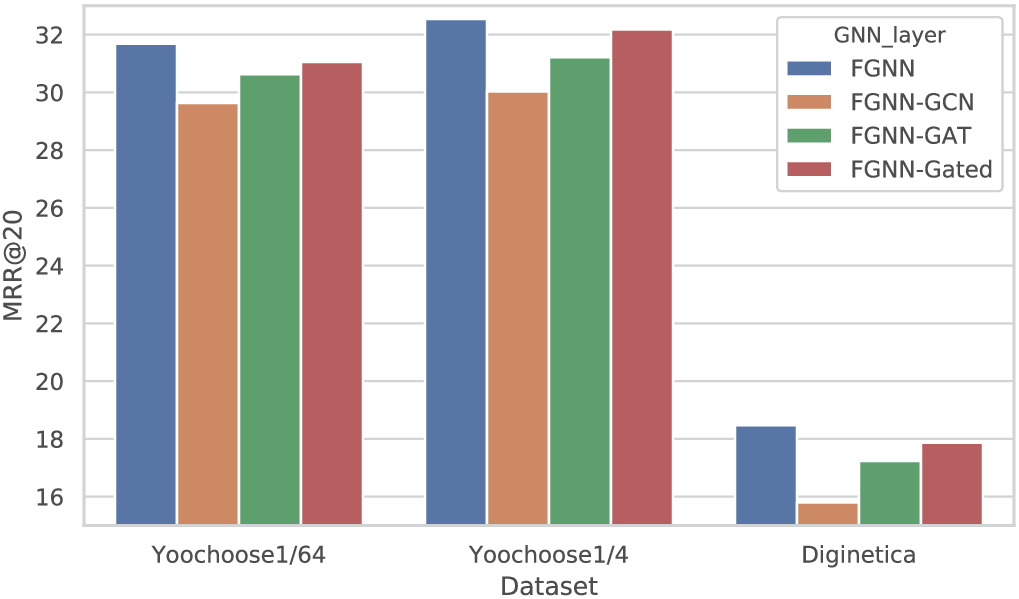}
    }
    \vspace{-0.5cm}
    \caption{Results with different GNN layers.}
\end{figure*}

\begin{figure}[!ht]
    \centering
    \includegraphics[width=\linewidth]{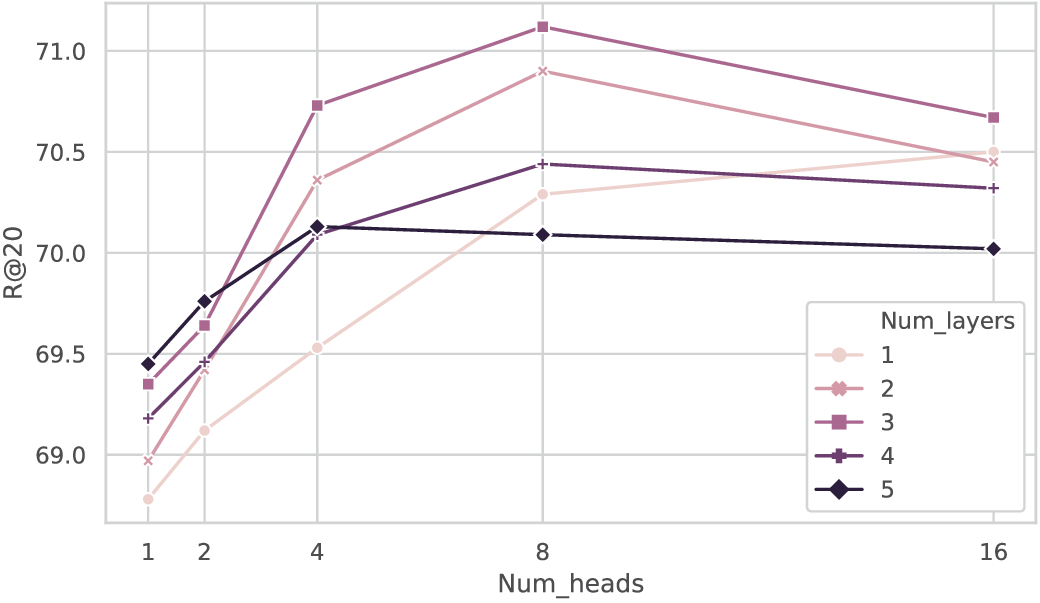}
    \vspace{-0.5cm}
    \caption{R@20 index for different number of layers and heads for WGAT.}
    \label{fig:p20-wgat-line}
\end{figure}

Different from the traditional methods mentioned above, all baselines using neural networks achieve a large performance margin. GRU4REC is the first to apply RNN-like units to encode the session sequence. It sets the baseline of neural-network-based methods. Though RNN is perfectly matched for sequence modeling, session-based recommendation problems are not merely a sequence modeling task because the user's preference is even changing within the session. RNN takes every input item equally importantly, which introduces bias to the model during training. For the subsequent methods, NARM and STAMP, which both incorporate a self-attention over the last input item of a session, they both outperform GRU4REC in a large margin. They both use the last input item as the representation of short-term user interest. It proves that assigning different attention on different inputs is a more accurate modeling method for session encoding. Looking into the comparison between NARM, combining RNN and attention mechanism, and STAMP, the complete attention setting, there is a conspicuous gap of performance that STAMP outperforms NARM. This further demonstrates that directly using RNN to encode the session sequence can inevitably introduce bias to the model, which the attention can completely avoid.

SR-GNN uses a session graph to represent the session sequence, followed by a gated graph layer to encode items. In the final stage, it again uses a self-attention the same as STAMP to output a session embedding. It achieves the best result compared to all methods mentioned above. The graph structure is shown to be more suitable than the sequence structure, the RNN modeling, or a set structure, the attention modeling.

\subsubsection{Higher Standard Recommendation with $K=5,10$}
\label{sec:high-std}
For more detailed results in Table~\ref{tab:5and10}, FGNN also achieves the best results with a higher standard of the top-5 and top-10 recommendation. The proposed FGNN model outperforms all baseline methods above. It has a more accurate node-level encoding tool, WGAT, to learn more representative features and a Readout function, to learn an inherent order of nodes in the graph to avoid the entire random order of items. According to the result, it is demonstrated that a more accurate session embedding is obtained by FGNN to make effective recommendations, which proves the efficacy of the proposed FGNN.

\subsection{Comparison with Other GNN Layers (RQ2)}
\label{rq2}
To efficiently convey information between items in a session graph, we propose to use WGAT, which suits the situation of the session better. As mentioned above, there are many different GNN layers that can be used to generate node embeddings, e.g., GCN~\cite{kipf2017semi}, GAT~\cite{velickovic2018graph} and gated graph networks~\cite{li2015gated,wu2018session}. To prove the usefulness of WGAT, we substitute all three WGAT layers with GCN, GAT and gated graph networks respectively in our model. For GCN and GAT, they both initially work for unweighted and undirected graph, which is not the same setting as the proposed session graph. To make both of them work on the session graph, we directly convert the session graph into undirected by replacing the original directed edges with undirected ones, i.e., reverse the source node and target node of edges. And we simply omit the original weight of edges and set all connections between nodes with the same weight 1.For the other one, the Gated graph networks, it can work with the session graph setting in its original form without any modification on the session graph.

In Figure~\ref{fig:p20-wgat} and Figure~\ref{fig:mrr20-wgat}, results of different GNN layers are shown with R@20 and MRR@20 indices. FGNN is the model proposed in this work, which achieves the best performance. WGAT is more powerful than other GNN layers in session-based recommendation. GCN and GAT are not able to capture the direction and the explicit weight of edges, resulting in performing worse than WGAT and gated graph networks, which holds the ability to capture these information. Between WGAT and gated graph networks, WGAT performs better because of the stronger ability of representation learning.

For the study of WGAT, we test how the number of layers and heads affect the R@20 index performance on \textit{Yooshoose1/64}. In Figure~\ref{fig:p20-wgat-line}, we report the experiment results of different number of layers ranging in $\{1,2,3,4,5\}$ and heads ranging in $\{1,2,4,8,16\}$. It shows that stacking three WGAT layers with eight heads performs the best. Lower results are shown for smaller models for the reason that the capacity of them is too low to represent the complexity of the item transition pattern. According to the tendency of results for larger models, it shows that it is difficult to train these models and the overfitting is harmful to the final performance.

\begin{figure*}
    \centering
    \subfigure[R@20 index.]{
    \label{fig:p20-readout}
    \includegraphics[width=0.48\linewidth]{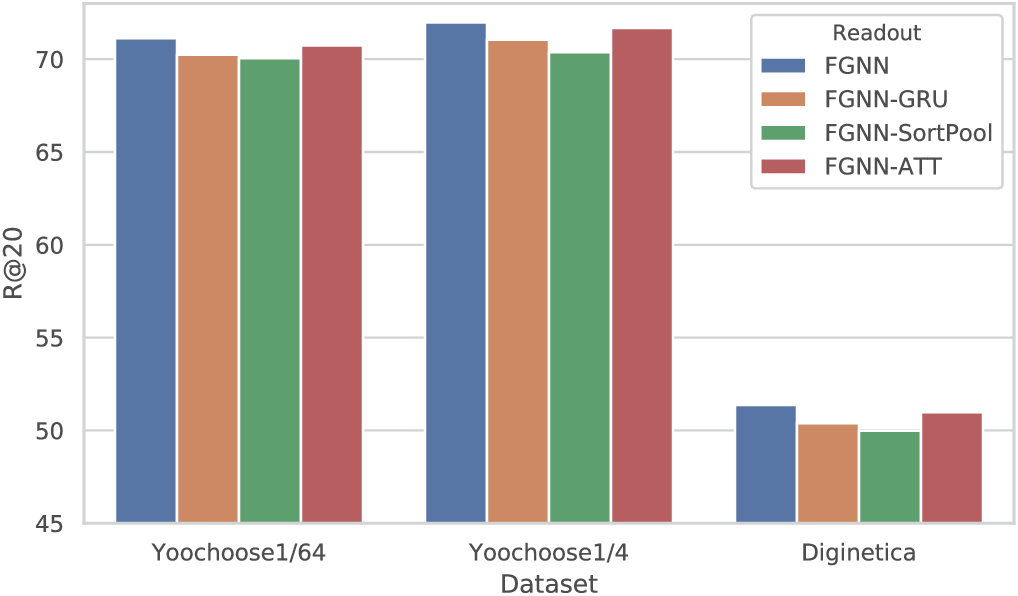}
    }
    \subfigure[MRR@20 index.]{
    \label{fig:mrr20-readout}
    \includegraphics[width=0.48\linewidth]{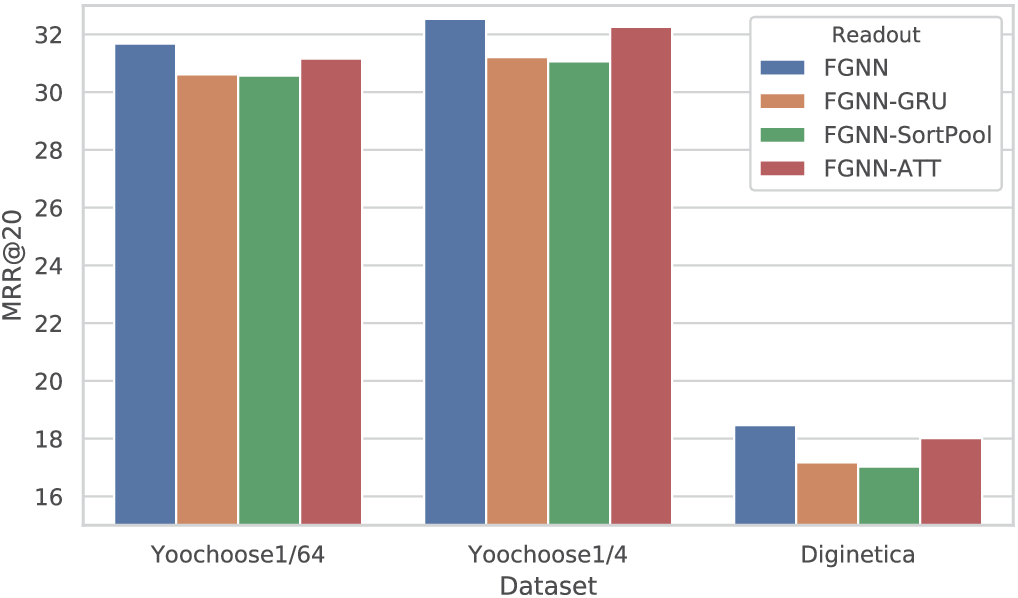}
    }
    \vspace{-0.5cm}
    \caption{Results with different aggregation functions.}
\end{figure*}

\begin{figure}[!ht]
    \centering
    \includegraphics[width=0.9\linewidth]{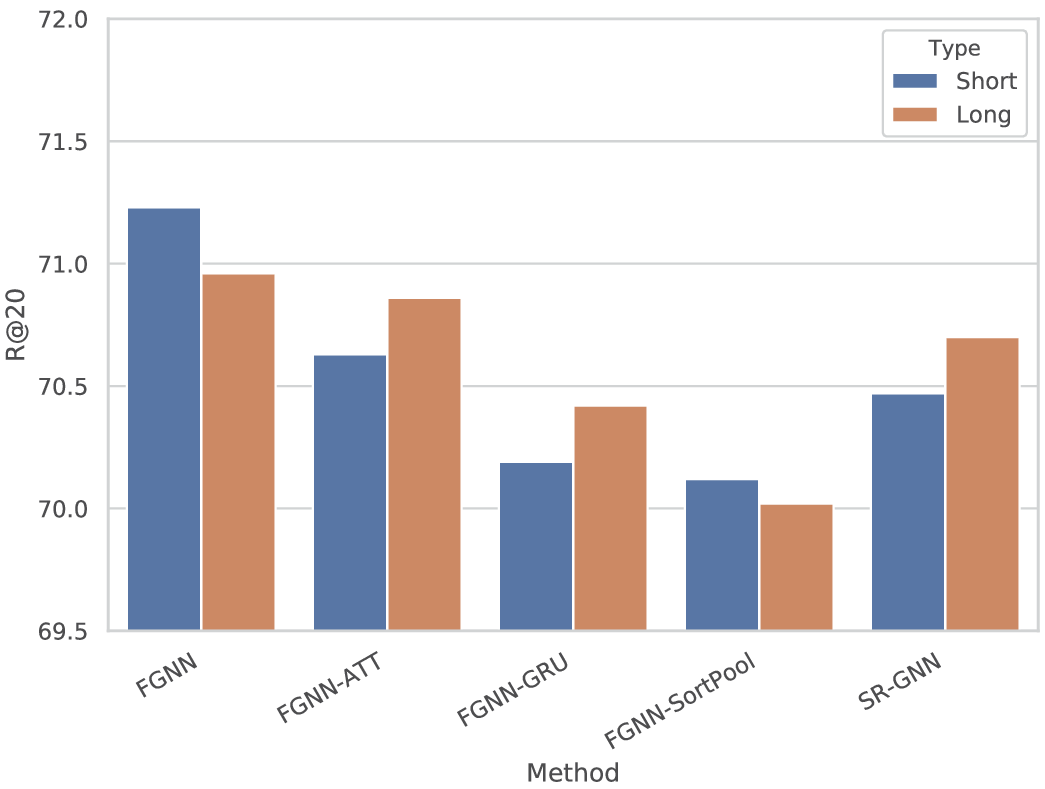}
    \vspace{-0.5cm}
    \caption{R@20 for Short and Long sessions with different aggregation functions and baselines.}
    \label{fig:p20-long-short}
\end{figure}

\subsection{Comparison with Other Graph Embedding Methods (RQ3)}
\label{rq3}
Different approaches for generating the session embedding after obtaining the node embedding stand for different emphasis of the input item. The Readout function proposed in this work learns an inherent order of the nodes by the query vector, which indicates the relatively different impact on the user's preference along with the item transition. To prove the superiority of our Readout function, we replace the Readout function with other session embedding generators:
\begin{itemize}
    \item \textbf{FGNN-ATT-OUT} We apply the widely-used self-attention of the last input item. It directly considers the last input item as the short-term reference and all other items as the long-term reference.
    \item \textbf{FGNN-GRU} To compare the inherent order learned by our Readout function, we use GRU to directly make use of the input session sequence order.
    \item \textbf{FGNN-SortPool} SortPooling is introduced by Zhang et. al.~\cite{zhang2018end} to perform a pooling on graph level by sorting the features of the nodes. This sorting can be viewed as a kind of order as well.
\end{itemize}

In Figure~\ref{fig:p20-readout} and Figure~\ref{fig:mrr20-readout}, results of different methods for graph level embedding generation are presented for all the datasets with the R@20 and MRR@20 indices. It is obvious that the proposed Readout function achieves the best result. For FGNN-GRU and FGNN-SortPool, they both contain an order but which is too simple to capture the item transition pattern. FGNN-GRU uses GRU to encode the session sequence with the input order. Such a setting is similar to the RNN-based method. As a consequence, it performs worse than attention-based method FGNN-ATT-OUT, which takes both the short-term and the long-term preference into consideration. As for FGNN-SortPool, it sorts the nodes based on WL colors from previous multiple layers of computations. Though it does not simply rely on the input order of the session sequence, the order used for the node is set according to the relative scale of the features. For the best performance, our Readout function learns the order of the item transition pattern, which is different from using the time order or the hand-crafted split of long-term and short-term preference. The results prove that there is a more accurate order for the model to make a more accurate recommendation.

For different session embedding generators, it is also important to look deep into how they perform on sessions with different lengths because the length varies greatly within one dataset. Following previous work~\cite{wu2018session,Liu18STAMP}, sessions in \textit{Yoochoose 1/64} are separated into two groups, i.e., \textbf{Short} and \textbf{Long}. \textbf{Short} indicates that the length of sessions is less than or equal to 5, while sessions longer than 5 are categorized as \textbf{Long}. Length 5 is the closest to the average length of total sessions. $70.1\%$ of \textit{Yoochoose1/64} are \textbf{Short} sessions and $29.9\%$ are \textbf{Long}. In addition to different session embedding generators, we take the previous GNN-based baseline method SR-GNN into comparison. For each method, we report the results evaluated in terms of R@20 in Figure~\ref{fig:p20-long-short}. In the aspect of both \textbf{Short} and \textbf{Long} sessions, FGNN achieves the best performance compared with other graph embedding generators and SR-GNN. The proposed Readout function shows superiority to other methods. The order introduced by the Readout function is demonstrated to convey more accurate information of item transition pattern. For the comparison among RNN-based and attention-based methods, it is shown that the performance is relatively better for longer sessions than shorter ones. This indicates that the item transition pattern relies on the latter input items of a sequence more heavily for long sessions, which is more suitable for these methods.

\section{Conclusion}
\label{conclusion}
Session-based recommender system is a challenging problem because the user history is unavailable for predicting the user's preference. This work proposes to use WGAT layers to learn item embeddings for items in a session, which are then processed by the Readout function to obtain the session embedding to represent the user's preference for this session. It is demonstrated by experiments that our proposed method achieves state-of-the-art results on benchmark e-commerce datasets. In the future, it is important and promising to make use of inter-session information to more accurately represent the user's preference.

\section{Acknowledgments} 
This work is supported by ARC DP190101985, DP170103954, NSFC 61628206 and NSFC 61806039.

\bibliographystyle{ACM-Reference-Format}
\bibliography{sample-base}

\end{document}